# Validation of Biometric Identification of Dairy Cows based on Udder NIR Images

Benjamin Schilling
Clarkson University
8th Clarkson Ave,
Potsdam, NY
schillbe@clarkson.edu

Keivan Bahmani
Clarkson University
8th Clarkson Ave,
Potsdam, NY
bahmank@clarkson.edu

Boyang Li
Clarkson University
8th Clarkson Ave,
Potsdam, NY
boyli@clarkson.edu

Sean Banerjee
Clarkson University
8th Clarkson Ave,
Potsdam, NY
sbanerje@clarkson.edu

Jessica Scillieri Smith
NYS Department of
Agriculture and Markets
Albany, NY
Jessica.ScillieriSmith
@agriculture.ny.gov

Tim Moshier
Acumen Detection
6274 Running Ridge Rd,
Syracuse, NY
tmoshier@acumendetection.com

Stephanie Schuckers
Clarkson University
8th Clarkson Ave,
Potsdam, NY
sschucke@clarkson.edu

## Abstract

*Identifying dairy cows with infections such as mastitis or cows on medications is an extremely important task and legally required by the FDA's Pasteurized Milk Ordinance. The milk produced by these dairy cows cannot be allowed to mix with the milk from healthy cows or it risks contaminating the entire bulk tank or milk truck. Ear tags, ankle bands, RFID tags and even iris patterns are some of the identification methods currently used in the dairy farms. In this work we propose the use of NIR images of cow's mammary glands as a novel biometric identification modality. Two datasets, containing 302 samples from 151 cows has been collected and various machine learning techniques applied to demonstrate the viability of the proposed biometric modality. The results suggest promising identification accuracy for samples collected over consecutive days.*

## 1. Introduction

Sales of milk produced from dairy cows is a multi-billion-dollar industry [1]. It is crucial to ensure the quality of the milk as a healthy food product and maintain the trust of the consumer. The Pasteurized Milk Ordinance (PMO) prohibits the sale of visibly abnormal milk secondary to inflammation of the mammary gland due to infection or other causes [2]. In addition, for animals being treated with medications, the PMO obligates dairy farmers to withhold and discard the milk from animals being treated with medications during and even after the treatment for sufficient amount of time (withholding period of the drug) to ensure no residual medication is present in the milk before sale [2]. While in order to satisfy the PMO requirements, every load of milk is being checked for beta lactam antibiotic residues and other signs of abnormalities (somatic cell counting methods, bacteria counts [3]), the presence of medications that are not routinely being tested could still lead to various human health concerns such as antibiotic resistance [2], [4], [5].

Regardless, if the cow is lactating, which is typically for 305 days after parturition, she must be milked. Dairy farms often use Radio-Frequency Identification (RFID) or plastic ear tags with unique management numbers to keep track of the cows [6]. The RFID systems results in relatively accurate identification rate. However, they suffer from installation and operational challenges. These systems are susceptible to tampering and damage and might not be economically viable for small farms [7], [8].

Even in bigger farms that use RFID technology, after the initial identification during the cows' ingress into the milking parlor, cows may shuffle and no longer be in the order that they entered, resulting in incorrect identification at the milk meter. As a result, during the milking process, farmers must confirm cows identify to physically move cows with abnormal milk, therefore preventing unhealthy cows from being milked at the wrong time by having those being treated in a separate area to prevent co-mingling their milk with the normal milk. Additionally, due to the PMO restrictions on Grade "A" milk, dairy farmers must visually check the milk for any abnormality (color, odor or texture) after the milking process [2]. However, at this point in the milking process the orientation of the cow in the milking stall may result in the ear tags not being visible to the milking staff, making identification via ear tag of cows with abnormal milk difficult.

In this work, we investigate the possibility of employing (Near-Infrared) NIR imaging technology [9] to facilitate accurate identification process of the dairy cows based on the images of the cows' mammary glands. The mammary gland complex has the advantage of being located where milking occurs, is easily accessible to milking staff and there is a possibility of integrating the NIR imaging hardware with the milking equipment "claw". The NIR imaging process can be employed in a cooperative and non-intrusive manner. This results in a cost-effective identification method which can potentially decrease the risk of disturbing the cattle, increase the throughput of the parlor and can conveniently be integrated in to the claw as an on-farm identification method. Additionally, the low cost and non-intrusive nature of this technology makes it a good candidate for fusing with other identification technologies employed in the parlor in order to improve the overall system's accuracy.

Using biometric traits has gained a lot of attention in current animal identification systems [7], [10], [11]. Previous work on animal biometrics investigated biometric methods such as retina [12], muzzle [13], [14] and iris [15]. However, these methods are costly, intrusive and in some cases may not be accurate [15]. In this work we propose the use of NIR images of dairy cows' mammary glands to identify and verify dairy cows' identity. The goal of this work is to introduce a novel biometric trait based on the NIR images of cows' mammary glands and investigate the viability and limitations of this modality in identification of dairy cows. To the best of our knowledge there is no publicly available dataset for NIR images of dairy cows' mammary glands. As a result, two datasets containing 302 images of 151 cows have been collected. We extracted features from the mammary vein patterns as well as the characteristics and geometric location of the cow's teats and employed various machine learning algorithms to evaluate the accuracy of the proposed biometric modality.

The rest of the paper is as follows: Section 2 describes the design of the equipment used in the process of creating the datasets as well as the preprocessing stages in the preparation of the dataset. In Section 3, we discuss the feature extraction process, while Section 4 presents the identification accuracy of various machine learning algorithms using the extracted features. Finally, Section 5 presents our conclusion and future work.

## 2. Dataset

In this section we discuss the equipment used for collecting the NIR images of the cow's mammary glands, as well as the processes involved with the collecting and preparation of the resulting datasets.

### 2.1. Equipment

We used a custom-built camera set up to capture the NIR images. The rig consisted of a 4-foot piece of metal chassis with wheels attached to the bottom. A Go Pro camera with NIR capability and two NIR lights were placed on top of the rig. Each IR light contains 68 LEDs with total power consumption of 500mA and range of 60 feet. Lastly, two 6000mAh external battery packs were used to power the IR lights. Figure 1. illustrates the top down view of the hardware used in this work.

This set up allowed us to capture 3840 by 2160 pixel images of the cows' mammary glands at 24 Frames Per Second (FPS). We captured the videos by placing the camera underneath the cows before the milking process. The study is under Animal Care and Use Committee (ACUC) protocol.

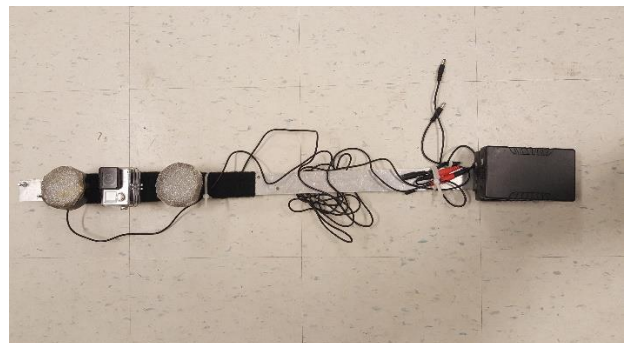

Figure 1: NIR Camera setup

### 2.2. Collection

Two datasets were collected where the first dataset contains the pre-singed mammary glands while the second dataset contains images of post-singed udders. Singe-ing is a typical process used in the dairy industry for removing the hair from the mammary glands' skin in order promote improved mammary hygiene. We hoped that the singing process would results in a cleaner and better image of the underlying veins. Thus, we captured the images in the second dataset after the cows underwent routine singeing of hair from their mammary glands. Both datasets consisted of images of the same cows from two consecutive days. The pre-singed dataset contains 150 images of 75 cows, while the post-singed dataset follows the same pattern and contains 152 images of 76 cows. There was a four-month break between the data collection for the first and second dataset. During this break period some cows ended their lactation and moved out of the group and were replaced by new cows. However, 21 cows are included in both datasets.

### 2.3. Pre-Processing

The farm used for this study milked in a double 10 parlor with cows oriented in a herringbone pattern. During the milking process 20 cows entered the milking parlor, 10 on each side. Immediately after entry, and before teat disinfection and the attachment of the milking cups, short video clips were taken of the cows' mammary glands. Figure 2 shows the milking parlor selected for the data collection. Figure 3 illustrates the ventral view of the bovine mammary glands complex. The complex is made up of four glands (quarters) with four teats (one teat per gland). The front quarters are just caudal to the abdomen. The Median Suspensory Ligament separates the right quarters from the left quarters. The front and rear quarters are separated by fine connective membranes. After collecting the data, human coders manually investigated each video clip and selected the cleanest frame possible to be included in the dataset. In addition to manually selecting the best possible frame for each cow, we put in place several other pre-processing steps to ensure the quality of the resulting datasets.

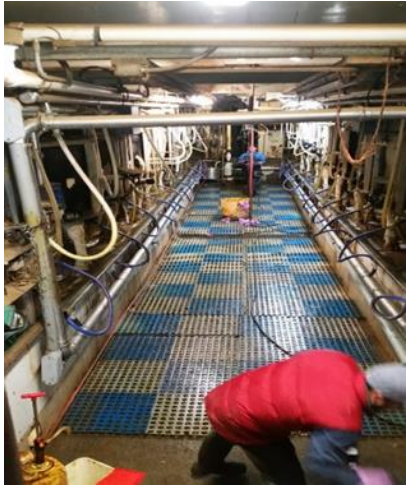

Figure 2 : Milking Parlor

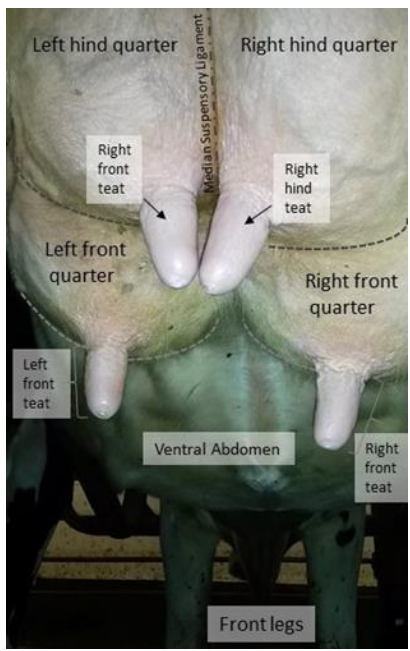

Figure 3: Ventral view of the bovine mammary glands complex as viewed by milking staff at the time of milking

Even though keeping the vertical distance between the cow's mammary gland and camera was straight forward as we filmed from the deck (floor of the parlor where the cows stand), keeping the filming angle due to the cows' placement in the parlor and horizontal distance (how far under the cow we should place the camera) turned out to be a very cumbersome task. To limit this issue, we started with a previously chosen quality frame, rotated the image so the cow is facing directly left. Then, we manually cropped the image around the mammary glands. The selected frame and the rotated and cropped frame can be seen in figures 4 and 5 respectively. This manual pre-processing would provide a region of interest for the feature extraction process and helps to remove the impact of any extraneous noise from the images. We repeated this process for every cow in the datasets.

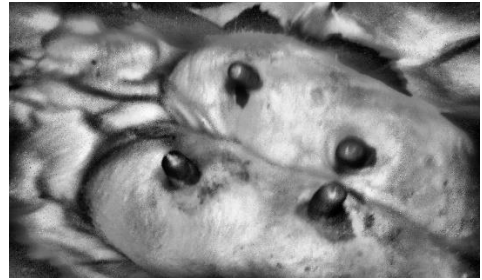

Figure 4 : Uncropped Frame

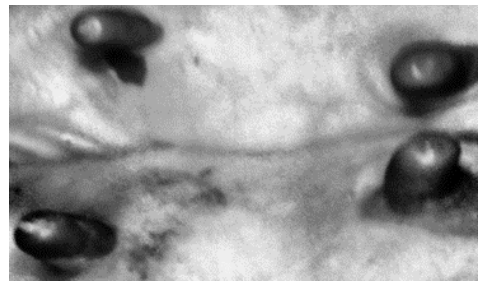

Figure 5 : Processed image

## 3. Feature Extraction and Matching

Starting with the rotated and cropped images, we observed the glands' vein pattern and superficial characteristics to find identifiable features.

### 3.1. Vein pattern

Palm and finger veins has been successfully used as a biometric modality [16], [17]. In this work, we explore the possibility of using local texture pattern descriptors as a feature extraction method for the mammary vein patterns. More specifically, we employed the rotationally invariant Local Binary Patterns (LBPs) [18]. Figure 6 represents an example of the LBP texture descriptor used in this work.

Starting with a grayscale image, for every pixel not on the border we compare it to the 8 surrounding neighbor pixels. This comparison has been presented with an 8-bit binary sequence. If the value of the center pixel is larger than neighbor pixel, we set the bit associated with that pixel to 1, otherwise the value of the bit will be set to 0. We used the 36 rotationally invariant combinations and radii of one and two pixels. Various window sizes have been explored however, the best result was yielded by the LBP extracted over the entire image. While investigating the mammary images for vein patterns, we encountered two major issues. Firstly, the presence of visible mammary veins was very inconsistent. Some cows had visible and distinctive vein patterns, while most had little to none. More study is needed to shed more light on the universality of vein patterns in dairy cows and its relationship to the cows' stage of lactation and age.

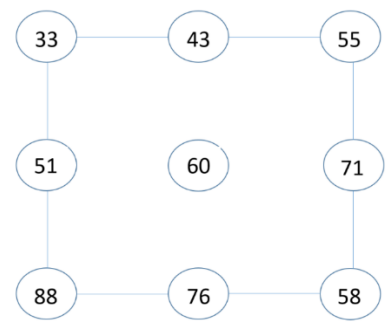

Figure 6: LBP operator

The second issue has to do with the environment where the cows are housed. Barns are typically messy, this can cause the cows to get mud and dirt on their mammary glands occluding the pattern beneath. The combination of these two factors caused vein patterns alone to be too inconsistent for identification.

### 3.2. Mammary glands

Another considerable feature of the cows' mammary gland is the four teats. The teats are consistent in cleanliness and unsusceptible to the environmental factors because they are not haired skin. Similar to the vein patterns, teats may change over the cow's lactation, predominantly in angle due to changes in milk production over time. However, we believe the day to day changes are not significant enough to disrupt the successful identification process. Additionally, cows suffering from mastitis or being treated may experience drastic decrease in the milk production which can potentially affect the angle and spacing of teats. Feature work can investigate the possibility of flagging sick cows based on these changes.

In this work, we extracted four different features from teats, namely; distance, interior angles, size and the aspect ratio of the surrounding box containing the mammary glands. We initially aimed to automatically place the bounding boxes using cascade object detection (Hough transforms) [19]. However, during our tests we observed that the accuracy of such methods was not acceptable for this task, likely due to the small nature of the dataset. Consequently, all of the frames in both datasets has been manually annotated by human coders. The manual annotation process allowed us to create datasets with higher quality as well as ensuring that any misidentification is solely due to the identification process and not a result of the low accuracy in the detection process of mammary glands. The manual annotation process used in this work is as follows: First, we calculate the pixel distances from the center of each teat to its 2 nearest neighbors forming a square (4 total distances). Then, we calculate the interior angles between each teat and others (4 total angles). Finally, we look at the characteristics of the teats themselves and calculate their aspect ratio and size. An example of an annotated from our datasets can be seen in Figure 7.

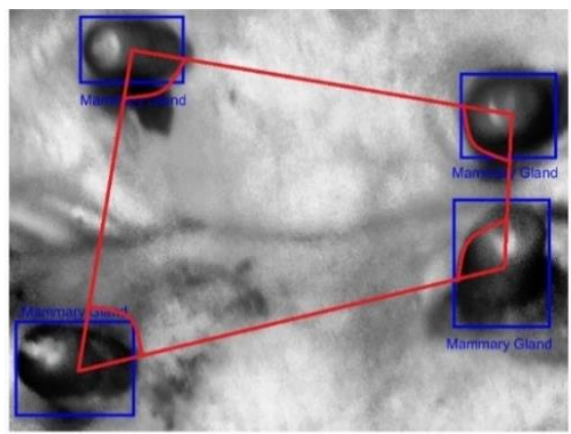

Figure 7 : Annotated NIR mammary image

## 4. Experimental Result

In addition to collecting and extracting the features, we also evaluated the performance of some of the most prominent machine learning algorithms on both datasets. We hope this analysis would provide readers with some insight on the viability of NIR imaging of cows' mammary glands and pave the way for the development of more sophisticated algorithms for this biometric modality. We employed the Scikit-learn machine learning library [20] and evaluated the identification accuracy of K-nearest Neighbors (KNN), Logistic Regression (LR), Support Vector Machine (SVM), Random Forest (RF) and Decision Tree (DT) classifiers in a one to many identification processes. In our analysis, we used the first images of the cows as a training set while the second set of images from the next day has been used for testing. Figure 8 represents an example of a subject with only slight day to day changes in the samples (difference between training and testing samples). On the other hand, Figure 9 illustrates one of the subjects with drastic changes in the images acquired over two consecutive days. Figure 10 and Figure 11 represent the identification accuracy of the evaluated classifiers over the combination of both datasets. Subjects were selected randomly and the reported identification accuracies are an average of 50 trials. It can be seen that using only the geometric features extracted from the teats resulted in higher identification accuracy across the range compared to geometric plus texture features of the veins. We believe this might be due to the inconsistencies in observing the vein patterns secondary to environmental factors. These results motivated us to evaluate each dataset using only the geometric features extracted from the teats. More exploration is needed to develop features which quantify vein and other shape/textual information of the mammary glands in future research. The remainder of the paper will only consider the geometric features of the teats.

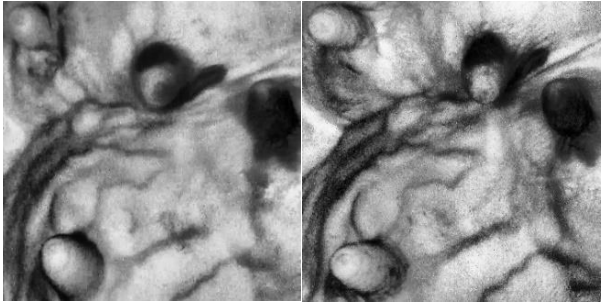

Figure 8: Example of a subject with persistent samples. Day 1 (left) to Day 2 (right)

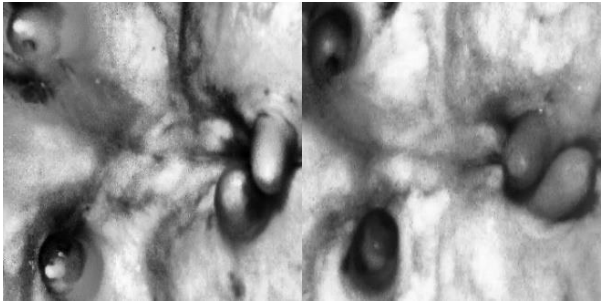

Figure 9: Example of a drastic changes in the samples captured in Day 1 (left) and Day 2 (right)

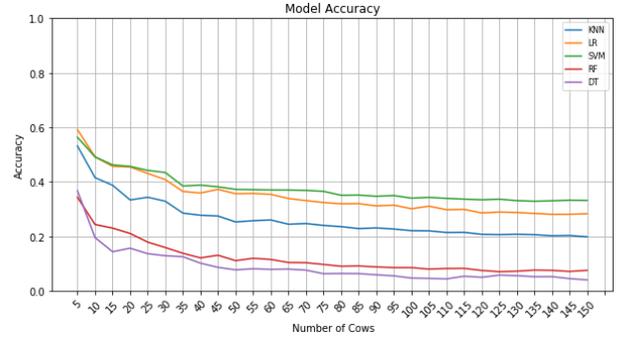

Figure 10: Identification accuracy over the combined dataset using vein patterns and mammary glands features

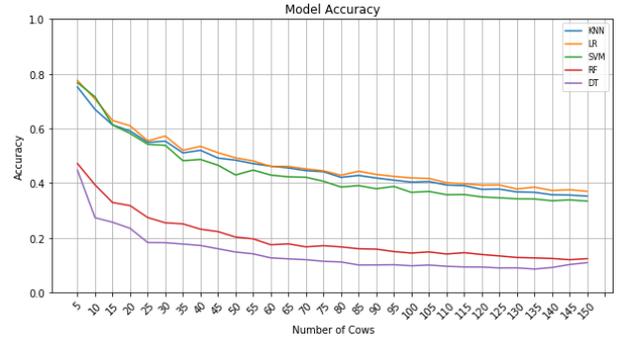

Figure 11: Identification accuracy over the combined dataset using mammary glands features

Figures Figure 12 and Figure 13 respectively represent the identification accuracy of the evaluated classifiers on the pre and post-singed datasets, using only the geometric features extracted from the teats. It is worth mentioning that many small milking parlors in US hold less than 20 cows at the same time during the milking process. Additionally, even in larger parlors, cows rarely switch places more than 1 or 2 away from they were registered when entering the parlor (error due to misreading in the RFID system). As a result, we mostly only need to identify less than 20 cows at a time.

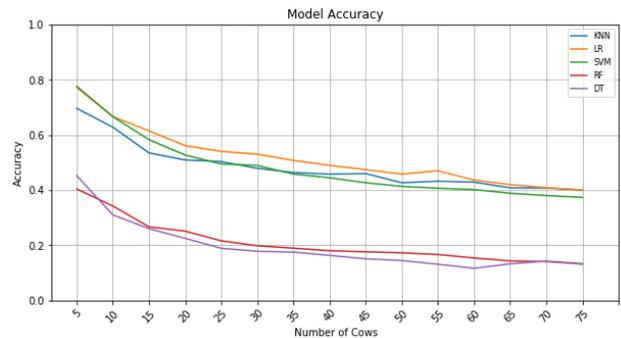

Figure 12: Identification accuracy over the pre-singed dataset using mammary glands features

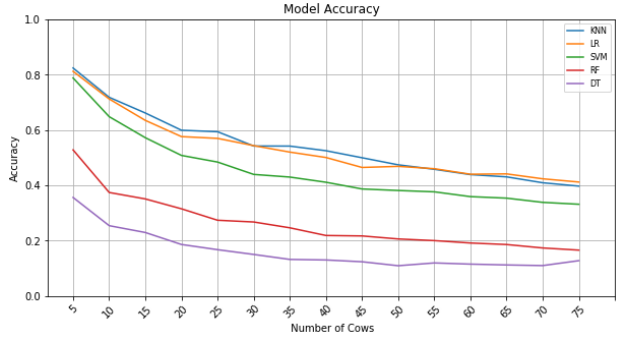

Figure 13: Identification accuracy over the post-singed dataset using mammary glands features

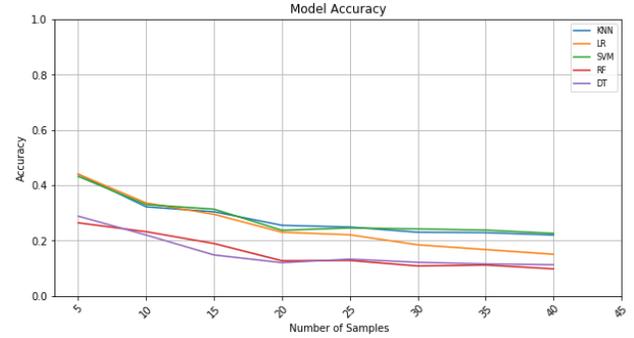

Figure 15: Identification accuracy over sample collected over 4 months

As mentioned in Section 2.2, twenty-one cows were seen in both data sets. An example of a cow in both datasets can be seen in Figure 14. Having the same cows in both datasets and observed over time, allowed us to also conduct a preliminary analysis on the permanence of the proposed biometric modality. We combined the repeated cows from the first (pre-stringed) and second (post-stringed) datasets into the third dataset of 42 cows. The samples from the first collection were used for the training process, while the samples from the second collection (collected four months later) has been used as a testing set. Figure 15 illustrates the identification accuracy of the evaluated classifiers for random number of cows selected from the resulting third dataset.

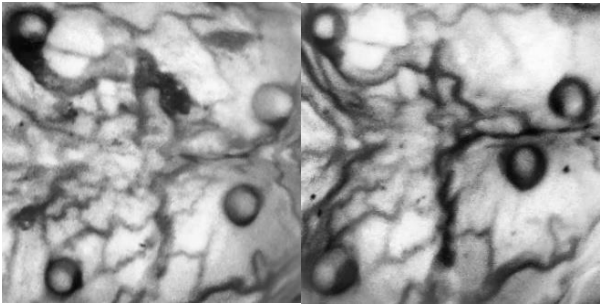

Figure 14: a subject in both dataset 1 (left) and dataset 2 (right)

The results presented in Figure 12 and Figure 15 reveals that even though the samples collected on consecutive days are consistent, over time, the changes in the mammary glands due to physiological changes over the cow's lactation can drastically affect the extracted features. These temporal changes over the cows' lactation can hinder the identification process. As a result, any algorithm developed based on the features extracted from the teats would benefit greatly from a rolling enrolment mechanism.

## 5. Conclusion and Future Work

This work introduces a novel biometric modality based on the NIR imaging of cows' mammary glands. The proposed method is non-intrusive and make use of inexpensive sensors. We produced the first publicly available datasets for NIR images of dairy cows' mammary glands and evaluated the possibility of using the characteristics of the teats as identifying features. The result suggests that NIR images of the mammary glands has potential to be used as a biometric modality. An average of ~60% identification accuracy was achieved for groups of twenty cows.

We did not observe consistent vein patterns, while textual analysis using LBP did not show promise, additional image-based textual features should be explored to improve the performance and shed lights on the consistency and permanence of the mammary veins.

The permanence of the features extracted from teats has been evaluated and the results suggest that the features extracted from the teats of healthy dairy cows are persistent enough for the day to day identification. Additionally, future work can investigate the possibility of flagging sick cows based on the extracted features from the teats.

The result obtained in this work shows promising identification accuracy, future work could apply other classification, feature extraction methods to increase the accuracy. The possibility of biometric fusion can be investigated to increase the overall accuracy of the identification system. Finally, deep convolution neural networks can be employed to extract features from the region of interest and help avoid the cumbersome task of manually annotation of the dataset.

## 6. Acknowledgement

This material is based upon work supported by the Acumen Detection, Center for Identification Technology Research,

and the National Science Foundation under Grant No. 1650503.